\documentclass[times,number]{elsarticle}
\usepackage{amsmath}
\usepackage{amssymb}
\usepackage{graphics}
\usepackage{graphicx}
\usepackage{epsfig}
\usepackage{dcolumn}
\usepackage{bm}
\usepackage{lineno}
\begin{document}
\begin{frontmatter}

\title{Study of uniformity of characteristics over the surface for triple GEM detector}

\author[label1]{S. Chatterjee}
\author[label1]{S. Chakraborty}
\author[label1]{S.~Roy\corref{cor1}}
\ead{shreyaroy@jcbose.ac.in}
\author[label1]{S.~Biswas\corref{cor}}
\ead{saikat@jcbose.ac.in, saikat.ino@gmail.com, saikat.biswas@cern.ch}
%\thanks[label1]{}
%\corauth[cor1]{}
\author[label1]{S.~Das}
\author[label1]{S.~K.~Ghosh}
\author[label1]{S.~K.~Prasad}
\author[label1]{S.~Raha}

\cortext[cor]{Corresponding author}

\address[label1]{Department of Physics and Centre for Astroparticle Physics and Space Science
(CAPSS), Bose Institute, EN-80, Sector V, Kolkata-700091, India}

\begin{abstract}
Study of the uniformity of gain, energy resolution and count rate over active area of a triple GEM detector has been performed using a strong Fe$^{55}$ X-ray source with premixed gas of Argon and CO$_2$ in 70/30 ratio and conventional NIM electronics. The detail method of measurement and experimental results are presented in this article.
\end{abstract}
\end{frontmatter}

\section{Introduction}\label{intro}
%\vspace{-0.35cm} 

Triple GEM chambers \cite{FS97} will be used in CBM experiment \cite{CBM} at FAIR \cite{FAIR} for muon tracking (Muon Chamber: MUCH). Keeping this in mind a study of uniformity of gain, energy resolution and count rate over the surface has been carried out for a standard 10 cm $\times$ 10 cm triple GEM detector using Fe$^{55}$ X-ray source. Argon and CO$_2$  gas mixture in 70/30-volume ratio has been used for this study. A low noise charge sensitive preamplifier and conventional NIM electronics has been used. The above-mentioned characteristics for the GEM detector have been measured on a scheme of 5$\times$4 array in the central part of the active area. Uniformity study using a collimated source of low activity is reported in Ref~\cite{RP}. The novelty of the present work is that high radiation ($\sim$~300~kHz) is used for the measurement of gain and energy resolution and the radiation is not collimated to a point but a patch of $\sim$~50~mm$^2$ has been exposed.

%\vspace{-0.6cm} 
\section{Detector descriptions and experimental set-up}\label{setup}
\begin{figure}[htb!]
\begin{center}
\includegraphics[scale=0.55]{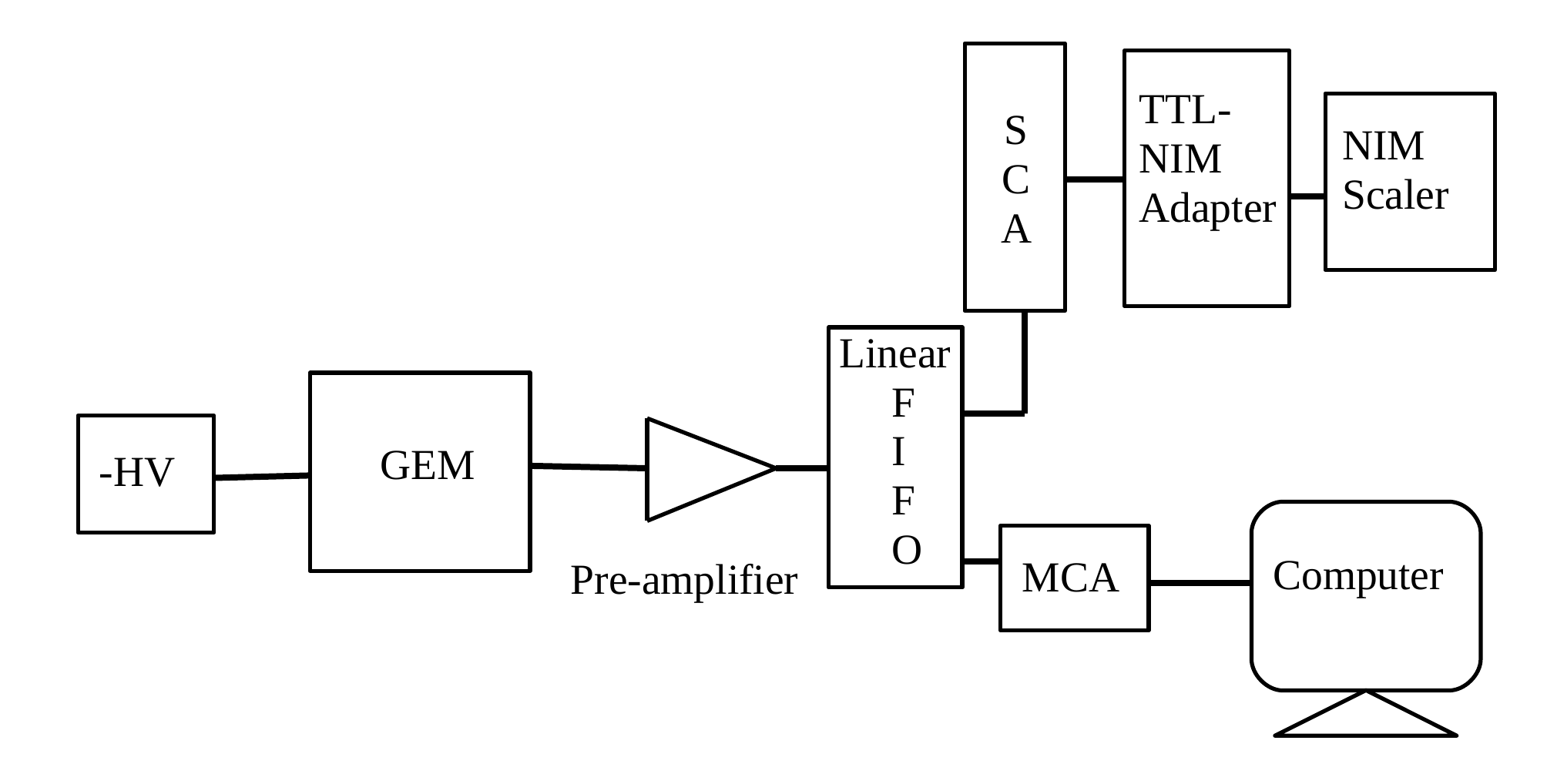}
%\vspace{-0.3cm} 
\caption{Schematic representation of the electronics setup.}
\label{block}
 \end{center}
\end{figure}

%\vspace{-0.4cm} 
In this study, a GEM detector prototype, consisting of three 10~cm~$\times$~10~cm double mask foils, obtained from CERN has been used. The drift, transfer and induction gaps of the detector are kept 3 mm, 2 mm and 2 mm respectively. The high voltages (HV) to the drift plane and individual GEM planes have been applied through a voltage dividing resistor chain. Although there is a segmented readout pads of size 9~mm~$\times$~9~mm each, the signal in this study was obtained from all the pads added by a sum up board and a single input has been fed to a charge sensitive preamplifier (VV50-2) \cite{Preamp}. The gain of the preamplifier is 2~mV/fC with a shaping time of 300~ns. A NIM based data acquisition system has been used after the preamplifier. Same signal from the preamplifier has been used to measure the rate and to obtain the energy spectrum. The output signal from the preamplifier has been fed to a linear Fan-in-Fan-out (linear FIFO) module for this purpose. The analog signal from the linear FIFO has been put to a Single Channel Analyser (SCA) to measure the rate of the incident particle. The SCA has been operated in integral mode and the lower level in the SCA has been used as the threshold to the signal. The threshold has been set at 0.1~V to reject the noise. The discriminated signal from the SCA, which is TTL in nature, has been put to a TTL-NIM adapter and the output NIM signal has been counted using a NIM scaler. The signal count rate of the detector in Hz is then calculated. Another output of the linear FIFO has been fed to a Multi Channel Analyser (MCA) to obtain the energy spectrum. A schematic representation of the set-up is shown in Figure \ref{block}.
%\vspace{-0.4cm} 

%%%%%%%%%%%%%%%%%%%%%%%%%%%%%%%%%%%%%%%%%%%%%%%%%%%%%%%%%%%%%%%%%%%
%%%%%%%%%%%%%%%%%%%%%%%%%%%%%%%%%%%%%%%%%%%%%%%%%%%%%%%%%%%%%%%%%%%

 %\vspace{-0.8cm} 

Pre-mixed Ar/CO$_2$ in 70/30 volume ratio has been used for the whole study. A constant gas flow rate of 3.4~litre/hour has been maintained using a V{\"o}gtlin gas flow meter. For all measurements a circular collimator of diameter 8~mm has been used to expose the X-ray from Fe$^{55}$ source.

%\vspace{-0.7cm} 
\section{Results}\label{res}
%\vspace{-0.4cm} 
In this study the gain, energy resolution and count rate are measured from the energy spectrum for the Fe$^{55}$ X-ray source. Initially these are measured varying the $\Delta$V as described in detail in Ref.~\cite{SRoy}. Since the experiment has been performed with a radiation source which emits a constant number of particles, a plateau in the count rate is reached at the highest efficiency of the chamber. The uniformity investigation of the detector described here has been carried out at an applied high voltage (HV) of -~4150~V corresponding to a $\Delta$V~$\sim$~385.9~V across each GEM foil. The active area of the chamber has been divided in 100 zones of 1 cm$^{2}$ and the above three parameters are measured in the central part in 5$\times$4 array i.e. in 20 zones.
%\vspace{-0.5cm} 

%%%%%%%%%%%%%%%%%%%%%%%%%%%%%%%%%%%%%%%%%%%%%%%%%%%%%%%%%%%%%%%%%%%
\begin{figure}[h!]
\begin{center}
\includegraphics[scale=0.3]{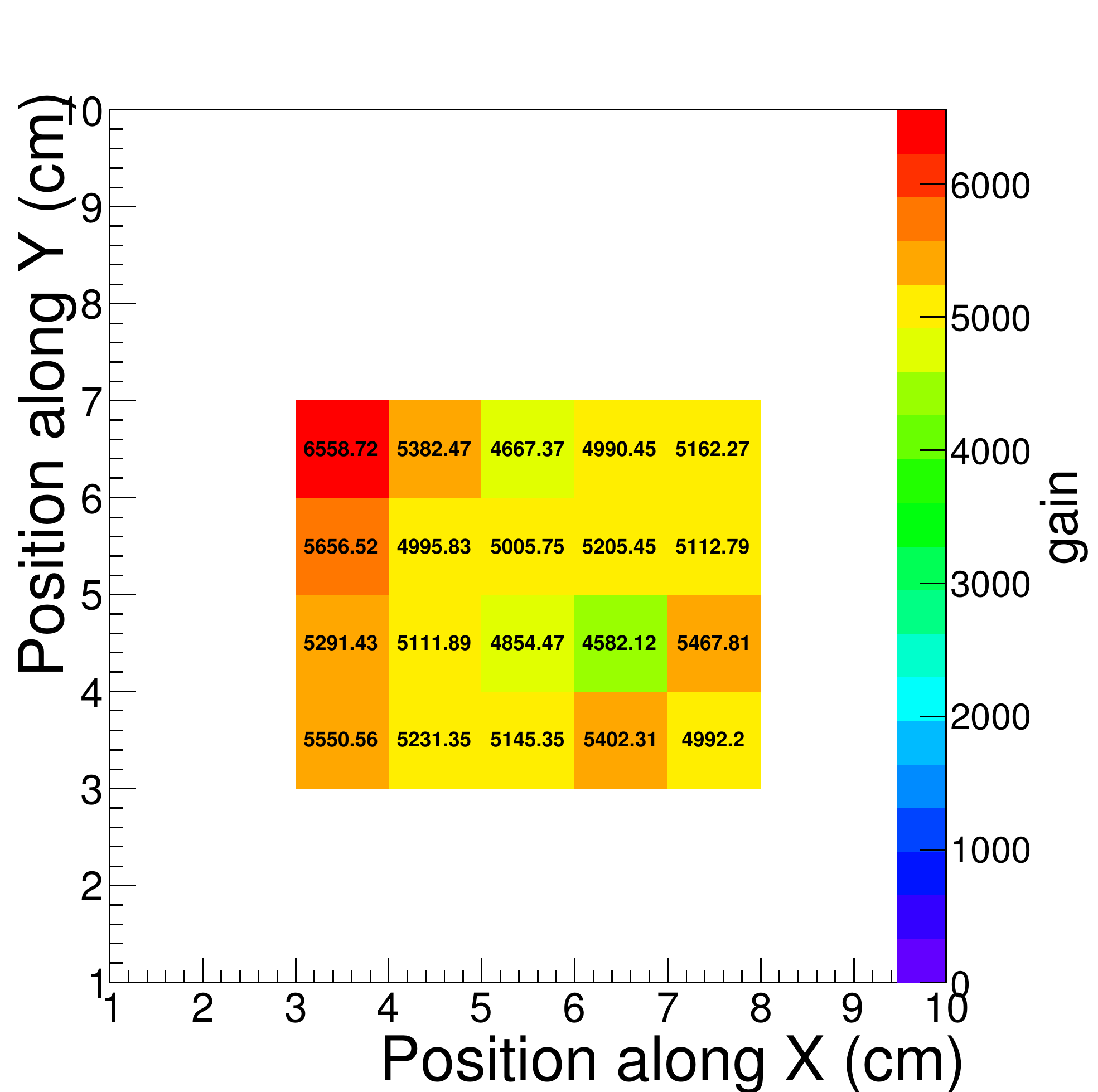}
\includegraphics[scale=0.3]{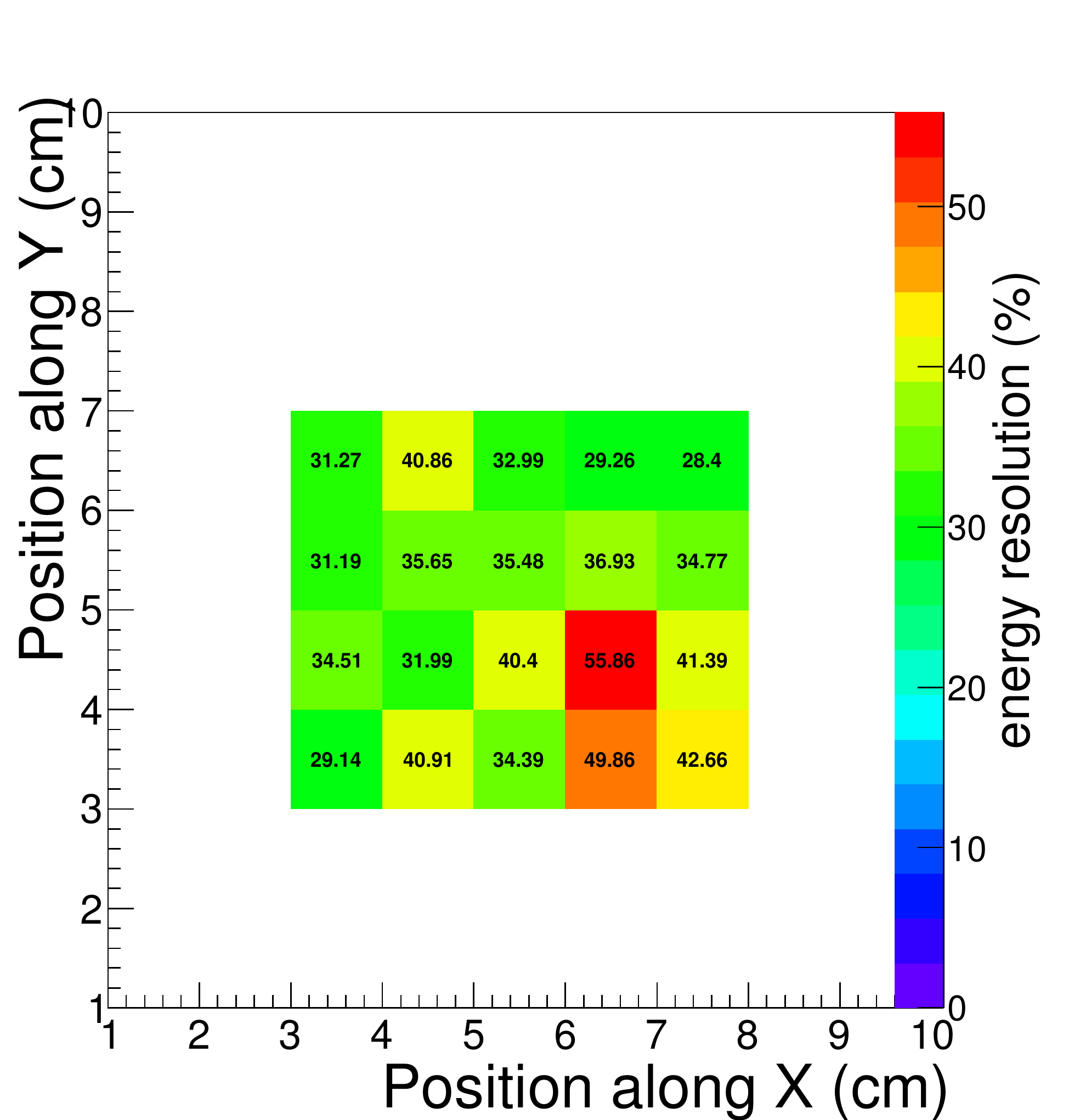}
\includegraphics[scale=0.3]{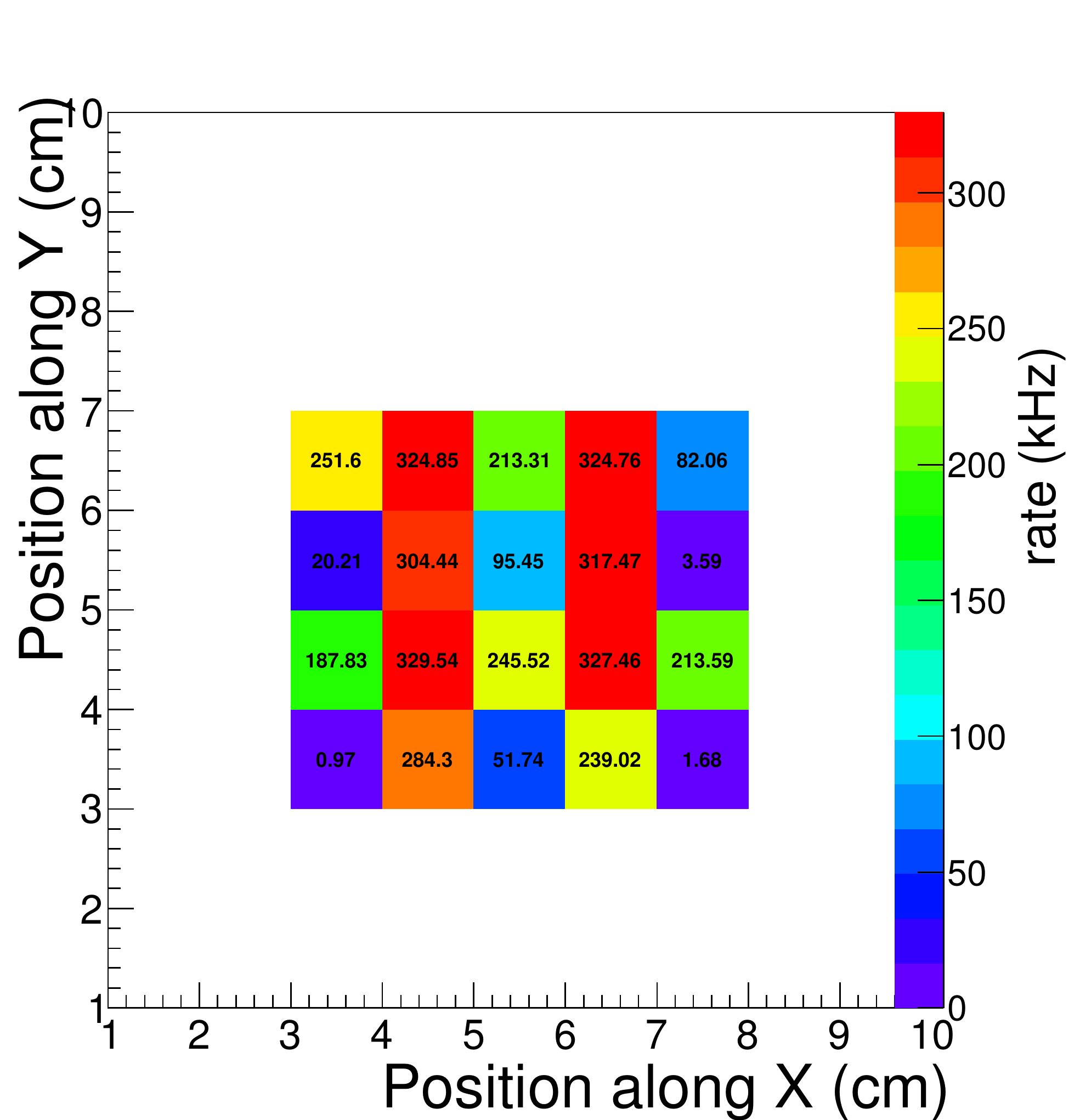}
\caption{Gain (top), energy resolution (middle), count rate (bottom) at 20 different places on the detector.}
\label{uniformity}
 \end{center}
\end{figure}
%%%%%%%%%%%%%%%%%%%%%%%%%%%%%%%%%%%%%%%%%%%%%%%%%%%%%%%%%%%%%%%%%%%
%\vspace{-0.7cm} 

Figure~\ref{uniformity} shows these three parameters measured at 20 different places at the central part of the detector. From Figure~\ref{uniformity} one can see that there is a non uniformity in the measured value of these parameters. 
%%\vspace{-0.1cm} 

%%\vspace{-0.7cm} 

\begin{figure}[h!]
\begin{center}
\includegraphics[scale=0.3]{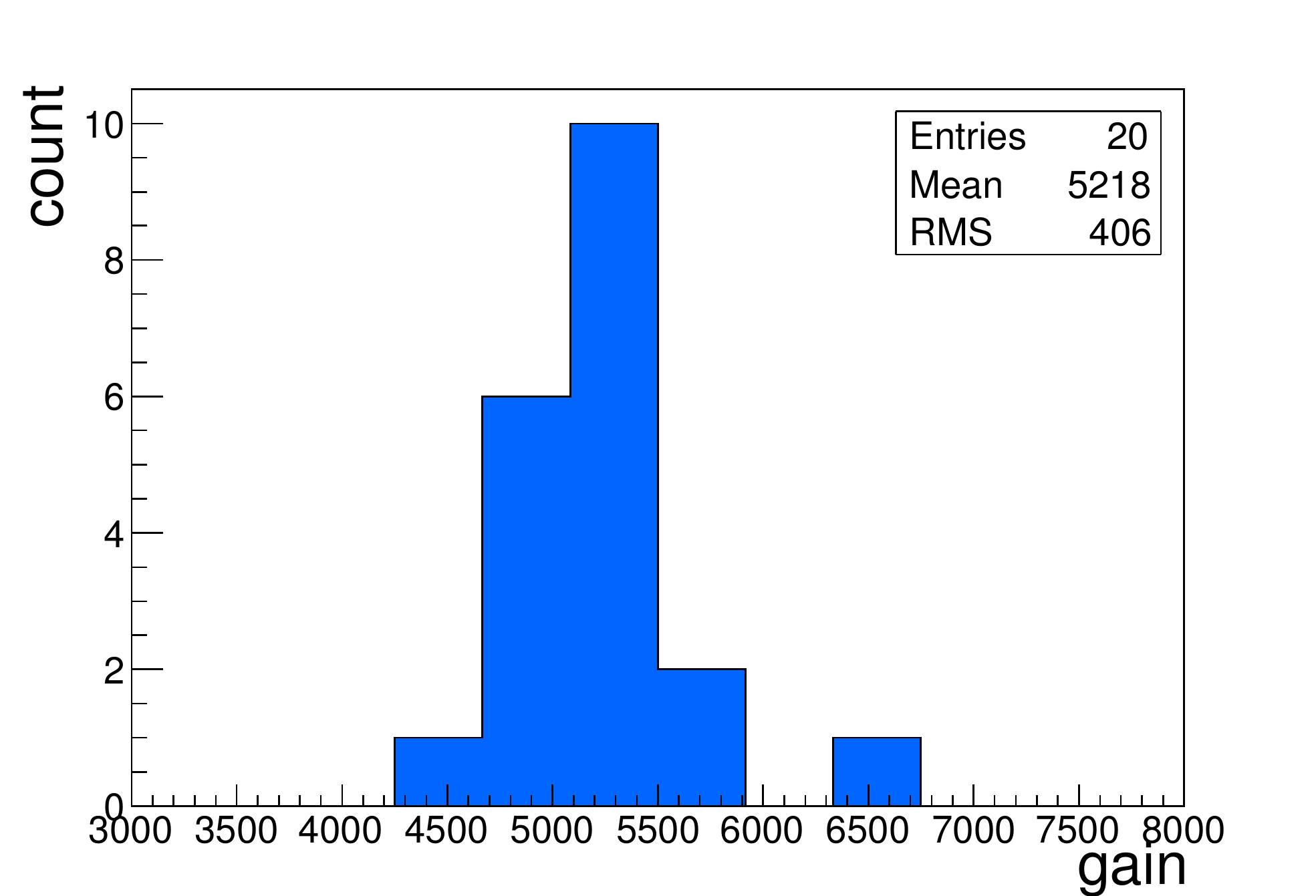}
\includegraphics[scale=0.3]{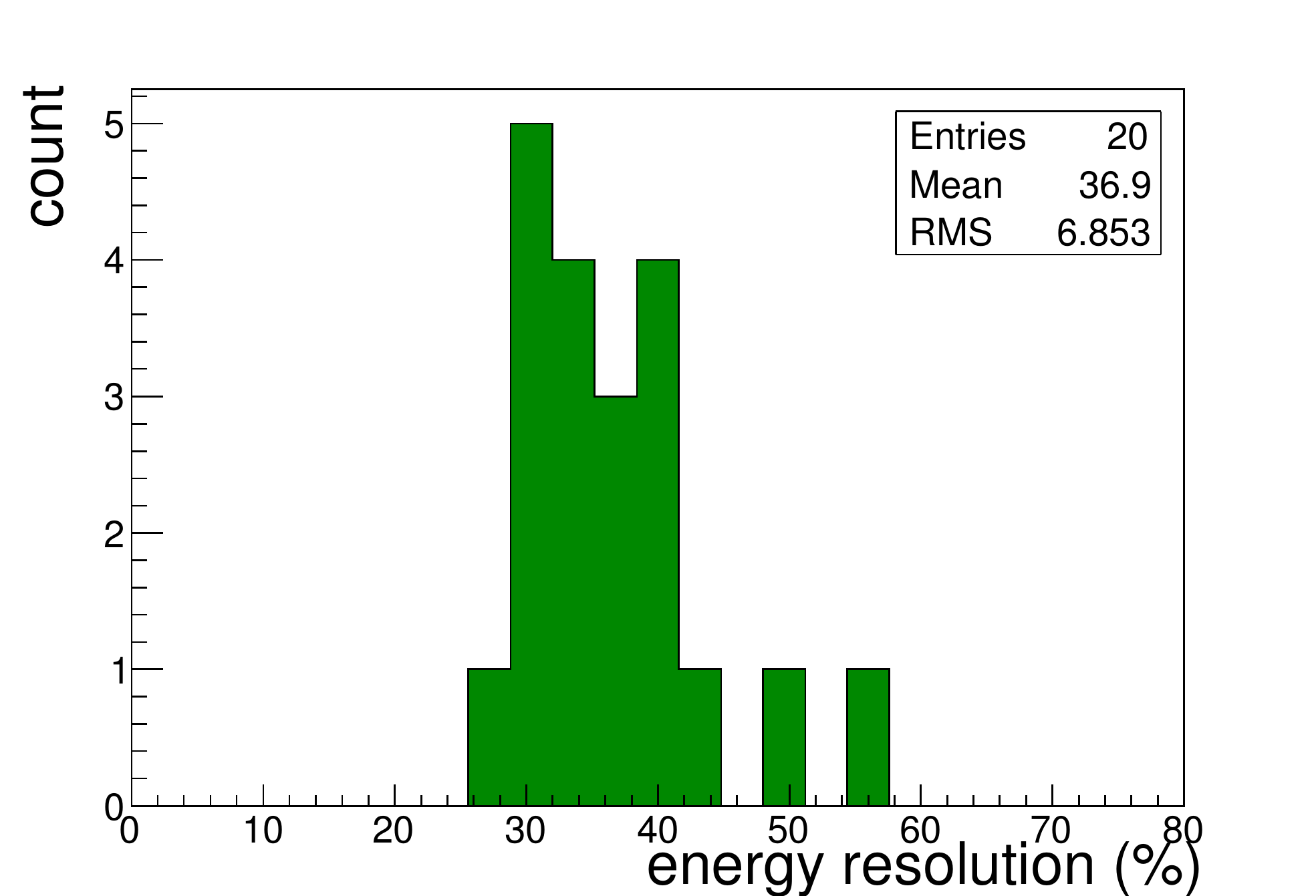}
\includegraphics[scale=0.3]{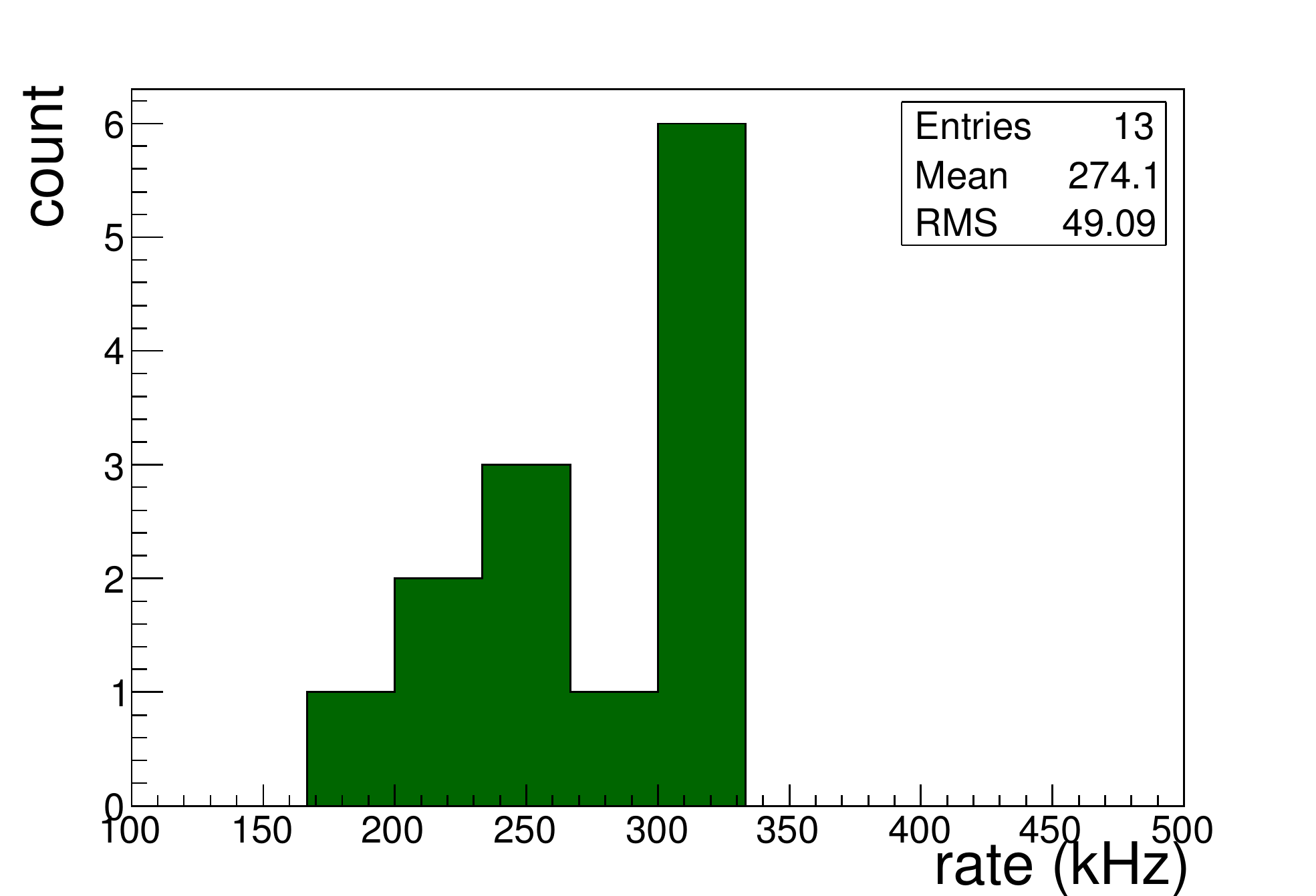}
\caption{Distribution of gain (top), energy resolution (middle), count rate (bottom).}
\label{histogram}
 \end{center}
\end{figure}

The distribution of these three parameters are shown in Figure~\ref{histogram}. For some zones the count rate has been found to be as low as 100~kHz. So for the distribution of the count rate a lower cut of 150~kHz has been used.
%%\vspace{-0.4cm} 

%\vspace{-0.5cm} 
\section{Conclusions}
%\vspace{-0.2cm} 

The characteristics of the GEM detector will not be the same over it's active area. It is to be mentioned here that because of intrinsic inhomogeneity in their characteristics due to GEM geometry variations and also for the inhomogeneity in the gap between individual GEM foils a gain variation up to few percentage is possible. In this study the gain, energy resolution and count rate have been measured at 20 places on the active area of the triple GEM detector prototype using a Fe$^{55}$ X-ray source to check the uniformity. For each measurement an area of $\sim$~50~mm$^2$ has been exposed by 5.9 keV X-ray. Over the measured area the gain fluctuation has been found to be $\sim$~10\% while the fluctuation of energy resolution and count rate is $\sim$~20\%.
%\vspace{-0.6cm} 

%%%%%%%%%%%%%%%%%%%%%%%%%%%%%%%%%%%%%%%%%%%%%%%%%%%%%%%%%%%%%%%%%%%
%%%%%%%%%%%%%%%%%%%%%%%%%%%%%%%%%%%%%%%%%%%%%%%%%%%%%%%%%%%%%%%%%%%
%\vspace{-0.2cm} 

%\vspace{-0.7cm} 
\section{Acknowledgements}
%\vspace{-0.2cm} 
The authors would like to thank Dr.~Christian~J.~Schmidt and Mr.~J{\"o}rg~Hehner of GSI Detector Laboratory for valuable discussions and suggestions in the course of the study and providing some components. This work is partially supported by the research grant SR/MF/PS-01/2014-BI from DST, Govt. of India and the research grant of CBM-MUCH project from BI-IFCC, DST, Govt. of India. S. Biswas acknowledges the support of DST-SERB Ramanujan Fellowship (D.O. No. SR/S2/RJN-02/2012).

%\vspace{-0.5cm} 

\end{document}